\documentclass[preprint2,6pt]{emulateapj}
\usepackage[T1]{fontenc}
\usepackage{lmodern}
\usepackage{graphicx}
\usepackage{amsmath}
\usepackage{amsfonts}
\usepackage{amssymb}%
\providecommand{\U}[1]{\protect\rule{.1in}{.1in}}

\newcommand{\figfactor}{1}
\newcommand{\nmfigfactor}{1}
\newcommand{\harmfigfactor}{1}

\newcommand{\R}{{\mathcal{R}}}
\newcommand*{\diff}{\mathop{}\!\mathrm{d}}


\begin{document}

\title{Jet Signatures in the Spectra of Accreting Black Holes}
\shorttitle{}
\author{Michael O' Riordan\altaffilmark{1}$^\dagger$, 
    Asaf Pe'er\altaffilmark{1}, 
    Jonathan C. McKinney\altaffilmark{2}}
\shortauthors{O' Riordan, Pe'er, \& McKinney}

\altaffiltext{1}{Physics Department, University College Cork, Cork, Ireland}
\altaffiltext{2}{Department of Physics and Joint Space-Science Institute,
    University of Maryland, College Park, MD 20742, USA}

\email{$^\dagger$michael\_oriordan@umail.ucc.ie}

\begin{abstract}

Jets are observed as radio emission in
active galactic nuclei and during the low/hard state in X-ray binaries 
(XRBs), but their contribution at higher frequencies has been uncertain. 
We study the dynamics of jets in XRBs using the general-relativistic 
magnetohydrodynamic code \texttt{HARM}. We calculate the high-energy
spectra and variability properties using a general-relativistic radiative 
transport code based on \texttt{grmonty}.
We find the following signatures of jet emission (i) a significant $\gamma$-ray 
peak above $\sim10^{22}\,\text{Hz}$, 
(ii) a break in the optical/UV spectrum, with a change from 
$\nu L_\nu \sim \nu^0$ to $\nu L_\nu \sim \nu$,
followed by another break at higher frequencies where the spectrum roughly 
returns to $\nu L_\nu \sim \nu^0$, and
(iii) a pronounced synchrotron peak near or below $\sim 10^{14}\,\text{Hz}$
indicates that a significant fraction of any observed X-ray emission originates 
in the jet.
We investigate the variability during a large-scale magnetic field inversion in
which the Blandford-Znajek (BZ) jet is quenched and a new 
transient hot reconnecting plasmoid is 
launched by the reconnecting field. The ratio of the $\gamma$-rays to X-rays 
changes from $L_\gamma / L_X > 1$ in the BZ jet to $L_\gamma / L_X < 1$ during
the launching of the transient plasmoid.

\end{abstract}
\maketitle

\section{Introduction}
Jets are observed in a wide range of accreting black hole systems, from 
stellar-mass black holes in X-ray binaries (XRBs), to supermassive black holes 
in active galactic nuclei (AGN). 
It is widely accepted that jets are responsible for the radio emission observed
both in AGN and during the low/hard state in XRBs 
\citep[see e.g.,][]{RM06,Fender10}, however the role of jets in producing
high-energy emission is still debated. 
In particular, there is no consensus 
regarding the origin of the X-ray component in the low/hard state in XRBs.
It has long been argued that inverse Compton emission from 
a corona of hot electrons surrounding the inner regions
of the disk can significantly contribute to the X-ray spectrum
\citep[e.g.,][]{Titarchuk94,MZ95,Gierlinski+97,Esin+97,Esin+01,Poutanen98,
    CB+06,Yuan+07,NM08,Niedzwiecki+14,Niedzwiecki+15,QL15}.
While X-rays are expected from the inner disk/corona, it is also 
possible that the X-rays are produced by jets
\citep[e.g.,][]{MiRo94,Markoff+01,Markoff+03,Markoff+05,FKM04,BRP06,Kaiser06,
    GBD06,Kylafis+08,Maitra+09,PC09,PM12,Markoff+15}.
This latter view has
largely been motivated by the observed correlation between the radio and X-rays
in the low/hard state \citep{Corbel+00,Corbel+03,Gallo+03}.  
The relative importance of
the disk and jet in generating the X-rays is still the subject of active
research.  Breaking this degeneracy is important for developing
an understanding of jets and of the disk-jet connection in XRBs and other
sources.

While most works assume that energetic electrons (leptons) are responsible 
for the observed emission, hadronic models 
\citep{MB92,Aharonian00,MP01,Mucke+03,Romero+03,Bosch-Ramon+05}, 
in which the protons are accelerated to ultrarelativistic energies, 
might also play a role in explaining the source of X-ray emission from 
these systems. Here, we limit our analysis to leptonic models in which the 
electrons are the primary radiators.
For a review of the features of both leptonic and hadronic models as applied 
to blazars, see e.g., \citet{Boettcher10,Boettcher+13}.

While radio observations provide a wealth of evidence for the existence of jets 
in AGN and XRBs, there is little direct evidence of the conditions required for
jets to form at all.
The fact that jets exist in such a wide range of systems has led to the
suggestion that their creation and dynamics should be governed by ingredients
common to these systems. Models of jet launching therefore involve accreting 
plasma, magnetic fields, and the extraction of rotational energy either from a 
black hole \citep{BZ77}, or from the accretion disk itself \citep{BP82}. 

\citet{Livio+99} argued that the Blandford-Znajek (BZ) mechanism will not 
operate efficiently 
in standard ``thin disks'' \citep{SS73,NT73} due to the fact that the magnetic 
flux at the horizon can not be significantly larger than that of the inner disk.
\citet{N+03} predicted that, if the accretion flow drags in a strong poloidal 
magnetic field to the black hole, the magnetic pressure will disrupt further 
axisymmetric accretion. 
They suggested that such a ``magnetically arrested disk'' (MAD) could be very 
efficient at converting the rest-mass energy of the fluid into heat, radiation, 
and mechanical/magnetic energy. 
Their MAD model relies on the key assumption that thin disks can drag magnetic 
fields to the horizon. Recent work by \citet{TNM11} showed that the 
BZ mechanism can efficiently power relativistic jets,
provided enough magnetic flux accumulates near the black hole.

A different class of accretion flow models, which readily advect magnetic 
fields towards the black hole, are the so-called 
``advection-dominated accretion flows'' (ADAFs) 
\citep{NY94,NY95a,NY95b,Abramowicz+95,NM08,YN14}.
\citet{Avara+15}, with the inclusion of results from \citet{MTB12},
showed that the BZ mechanism produces much more powerful jets
in MAD ADAFs than in MAD thin disks.

Radiatively inefficient accretion flows (RIAFs), by definition, are 
flows for which the cooling time of a fluid element is much longer than the 
time required for the fluid element to be accreted onto the black hole.
Radiatively inefficient ADAFs have been used extensively to model low 
luminosity systems such as the low/hard state in XRBs 
\citep[see e.g.,][]{NM08,YN14}.
Here, low luminosity means that 
$L \ll L_\text{Edd}$, where $L_\text{Edd}$ is the Eddington luminosity.
These flows are geometrically thick, optically thin, and collisionless. 
Due to the fact that the
electrons and ions are collisionally decoupled, they are likely to be at 
different temperatures, although the details of the electron thermodynamics in 
these systems are still being developed \citep{Ressler+15}.
In what follows, we deal exclusively with radiatively inefficient ADAFs 
and will simply refer to these as RIAFs.

The equations of general-relativistic magnetohydrodynamics (GRMHD) describe 
accreting systems in which the radiation is dynamically unimportant i.e. RIAFs.
In the past decade, global GRMHD simulations \citep{Gammie+03,MG04} have 
greatly improved our understanding of accretion physics and jet launching.
In particular, recent numerical simulations of MADs 
\citep{TNM11,MTB12} have demonstrated the launching of highly
efficient jets by the BZ mechanism; we will refer to these jets as ``BZ jets''. 
These simulations show that the BZ jet efficiency (defined as the 
energy extracted versus energy lost to the black hole) in MADs can be 
$> 100$ per cent. This means that more energy flows out of the black hole than 
flows in, which can only be achieved by extracting rotational energy from the 
black hole.

While these GRMHD simulations give much information about the fluid dynamics and
possible jet launching mechanisms, the results can not be directly tested by
comparing with observational data. To bridge this gap between theory and 
observations, in recent years, there has been wide interest in adding radiation 
to these simulations. Including radiation is necessary both for 
calculating the observational signatures, and for extending the simulations to 
regimes where the radiation becomes dynamically important i.e., where
$L \gtrsim 10^{-2} L_\text{Edd}$ \citep{Dibi+12}.

Broadly speaking, there are two main approaches to treating the 
radiation. The first involves evolving the radiation field self-consistently 
with the matter, and is mainly used to calculate the effects of radiation on the
fluid dynamics. This approach is employed in the general relativistic radiation 
magnetohydrodynamics codes \texttt{KORAL} \citep{Sadowski+13}, 
\texttt{HARMRAD} \citep{McKinney+14}, and \texttt{bhlight} \citep{Ryan+15}.
\texttt{KORAL} and \texttt{HARMRAD} treat the radiation as a 
separate fluid and close the fluid equations using the M1 closure 
\citep{Levermore84}, in which the radiation field is assumed to be isotropic in 
some frame (not necessarily the fluid frame). This approach is formally accurate
at high optical depths, however fails to capture the frequency dependence 
required for Compton scattering, and the angular dependence expected at lower
optical depths.
\texttt{bhlight} solves the GRMHD equations using a direct Monte Carlo solution 
of the radiative transport equation. This approach has the advantage that the
frequency and angular dependences of the radiation field can be included, 
however, since it involves tracking photons individually, it is limited to a 
regime in which radiative effects play a sub-dominant but non-negligible role on 
the dynamics. \texttt{bhlight} has been optimized for calculating the effects
of radiation on the dynamical evolution, and so the spectral resolution at low
and high frequencies (which have little effect on the dynamics) is limited.

The second method involves calculating the radiation field in a 
post-processing step, using the fluid data as input.
Examples of general-relativistic radiative transport codes which employ a
post-processing approach include \texttt{grmonty} \citep{Dolence+09}, 
\texttt{ASTRORAY} \citep{SH11}, \texttt{GRay} \citep{Chan+13}, 
and \texttt{HEROIC} \citep{Zhu+15,Narayan+15}.
Since the fluid data is supplied by an external code, the 
post-processing algorithms can be optimized for calculating spectra and images.
The disadvantage of this approach is that it is only applicable in regimes in 
which the radiation is dynamically unimportant.
These codes have been used by many authors to calculate the 
observational signatures of low luminosity systems in which the radiation 
pressure can be neglected
\citep[e.g.][]{Moscibrodzka+09,MF13,Moscibrodzka+14,Chan+09,Chan+15a,Chan+15b,
    Shcherbakov+12,SM13}.
These works mainly focussed on reproducing the spectra and variability 
properties of Sgr A*, and place important constraints on quantities such as the
black hole spin, proton-to-electron temperature ratio, and inclination angle. 
The constraints placed on the proton-to-electron temperature ratio could also 
be relevant for the low/hard state in XRBs.

We use a similar post-processing approach, with a radiative transport code 
based on the freely available \texttt{grmonty}.
Here, we are interested in identifying the observational signatures of jet 
emission in XRBs.
Since our goal is to study jets, we use GRMHD simulations of RIAFs, 
supplied by the \texttt{HARM} code, as input for our post-processing 
calculation.
We perform our radiative transport calculations for both MAD and non-MAD RIAFs, 
and find significant differences in the resulting spectra.
Furthermore, we make a distinction between jet and disk emission, and keep
track of whether or not photons had some interaction (emission or scattering)
with the jet before escaping the system. This allows us to determine the jet 
contribution to the spectrum, and identify unique observational signatures of 
jets.

Our paper is organised as follows. In Section \ref{sec:model} we briefly
describe our 3D GRMHD simulations and radiative transport code. In Section
\ref{sec:results} we present our results, showing the observational jet
signatures and variability properties of the jet and disk emission.  In Section
\ref{sec:discussion} we discuss our findings and summarize our main results.

\newpage

\section{Model}
\label{sec:model}
\subsection{GRMHD simulation}
\label{subsec:grmhd sim}
\begin{figure}
    \includegraphics[width=\harmfigfactor\linewidth]{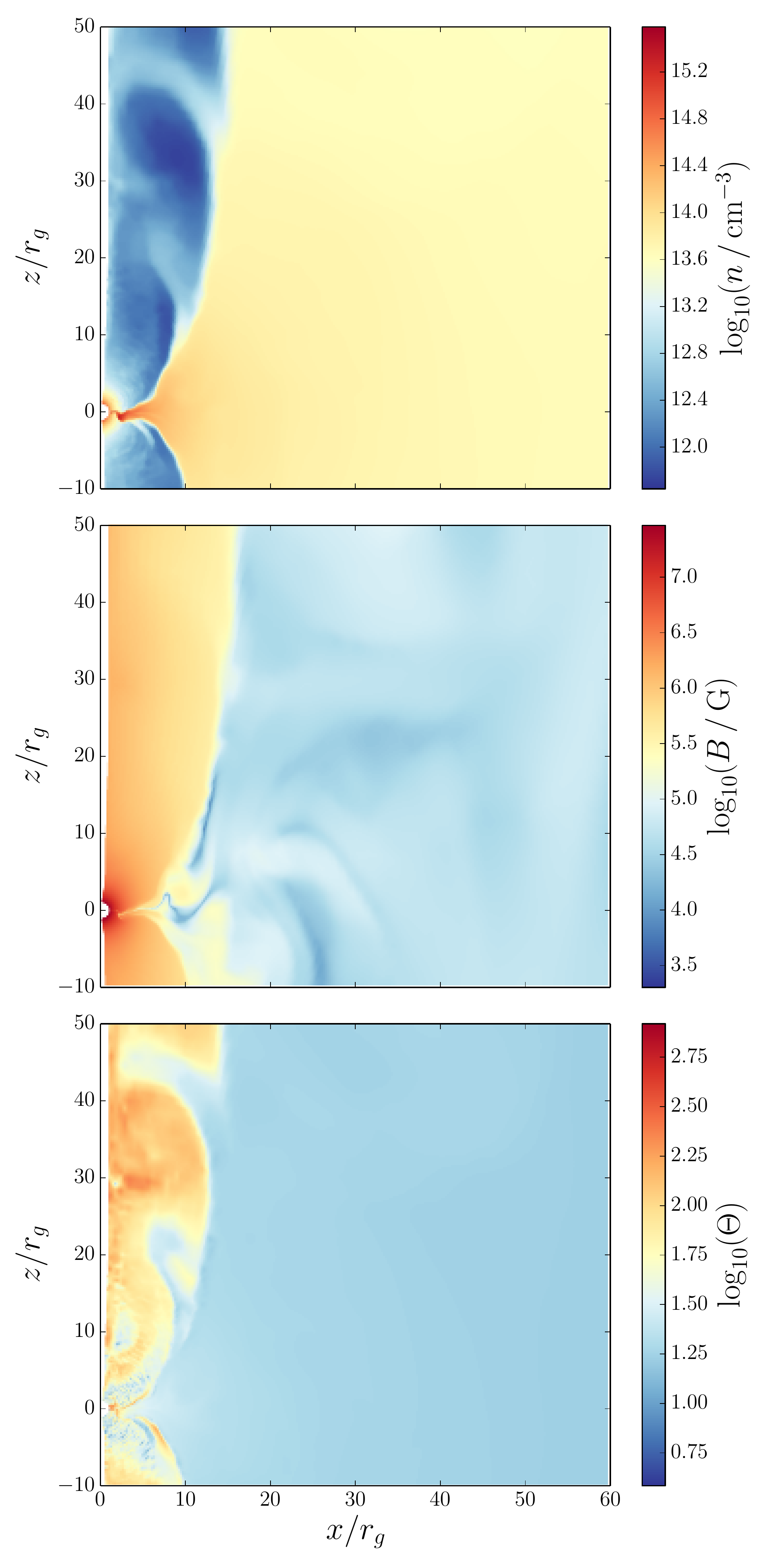}
    \caption{Electron number density, magnetic field strength, and 
        electron temperature, close to the 
        black hole, at $t = 26548 r_g/c$, in our MAD model.
        The inner $r \lesssim 10r_g$ of the disk is compressed by the
        black hole magnetosphere. The disk itself is geometrically thick, with
        approximately uniform density out to the boundary. The jet is visible as
        a lower density funnel region. The density enhancements in the jet are
        the result of QPOs driven by instabilities at the jet-disk interface.
        The jet region is highly magnetized, with
        $B \sim 10^6 - 10^7 \text{G}$.}
    \label{fig:harm_data_thickdisk7}
\end{figure}

We are interested in jets and so we focus on RIAFs, since these
are likely necessary for jet launching by the BZ mechanism 
\citep{Livio+99,Meier01,Avara+15}.
In this case, radiation is dynamically unimportant and the evolution is well 
described by standard GRMHD codes.
We use the \texttt{HARM} code \citep{Gammie+03,MG04}, which evolves the
GRMHD equations using a conservative, shock-capturing scheme. For our MAD model,
we choose the fiducial model, A0.94BfN40, from \citet{MTB12} in which the 
magnetic field has saturated near the black hole. In this magnetically choked 
accretion flow, the black hole magnetosphere compresses the inflow such 
that it becomes geometrically thin and the standard magneto-rotational 
instability is suppressed.  
The jet power in the BZ model is given by \citep{BZ77,TNM10,YN14}
\begin{equation}
    P_\text{BZ} = \frac{\kappa}{4\pi c} \Phi^2 \Omega_H^2
\end{equation}
where $\Phi$ is the magnetic flux threading the horizon,
$\Omega_H = a c / 2r_H$ is the angular velocity of the 
horizon, and $\kappa \approx 0.05$ is a dimensionless coefficient which depends 
weakly on the magnetic field geometry. The horizon radius, $r_H$, is given by 
$r_H = (1 + \sqrt{1 - a^2})r_g$, where $a$ is the dimensionless black-hole spin, 
$r_g = GM / c^2$, and $M$ is the mass of the black hole.
Thus, the highly magnetized state over most of the horizon
(see Figure \ref{fig:harm_data_thickdisk7}), 
and large black-hole spin ($a=0.9375$), 
are optimal for the BZ mechanism to generate powerful, relativistic jets 
\citep{TNM11,MTB12}. 

The initial mass distribution is an isentropic hydroequilibrium torus 
\citep{FM76,Gammie+03} 
with the inner edge at $r=10r_g$ and pressure maximum at $r=100r_g$. The
magnetic field has poloidal geometry with multiple loops of alternating polarity
for inducing magnetic field inversion/annihilation. These field 
inversions quench and relaunch magnetically dominated BZ jets 
(see Section \ref{subsubsec:magnetic field inversion}).

The jet forms as a highly magnetized, low density funnel
region along the spin axis of the black hole.
In Figure \ref{fig:harm_data_thickdisk7} we show snapshots of the 
electron number density $n$, magnitude of the magnetic field $B$, and 
dimensionless electron temperature $\Theta \equiv kT_e / mc^2$, 
at $t = 26548 r_g/c$.
These plots are scaled to the low/hard state in XRBs, with a 
black hole mass $M = 10M_\odot$ and accretion rate
$\dot M = 10^{-5} \dot{M}_\text{Edd}$, where $\dot{M}_\text{Edd}$ is
the Eddington accretion rate defined as $\dot{M}_\text{Edd} = L_\text{Edd} /
\left(0.1 c^2 \right)$ \citep[see e.g.,][]{NM08}.
The electron temperature shown corresponds to a proton-to-electron temperature 
ratio of $T_p/T_e = 10$ (see Section \ref{subsubsec:electron temperature}).
The inner $r \lesssim 10r_g$ of the disk is 
compressed by the black hole magnetosphere. The density enhancements in the jet
are due to instabilities at the jet-disk interface 
(see Section \ref{subsubsec:jet-disk qpos}). The horizon and funnel regions are both 
highly magnetized.
We use the ratio of the magnetic and rest-mass energy densities to define the 
jet, i.e. where $b^2 / \rho c^2 \geq \xi$, for some constant $\xi$. 
Here, $\rho$ is the rest mass density of the gas, and $b^2 = b^\mu b_\mu$, 
where $b^\mu$ is the magnetic field four-vector.
The precise value of $\xi$ is somewhat arbitrary and depends on the
particular simulation. We find that $\xi = 0.5$ gives a reasonable 
distinction between the jet and disk in our simulations.

It is also possible to distinguish between the jet, disk, and magnetized 
wind. The wind can be defined roughly by the condition that 
$b^2 / \rho c^2 < \xi$ and $\beta_p < 2$ \citep{MTB12}, where 
$\beta_p = p_\text{gas} / p_\text{mag}$ is the ratio of gas and magnetic
pressures. The disk then corresponds to the region with 
$b^2 / \rho c^2 < \xi$ and $\beta_p \geq 2$.
In our MAD simulation, the disk is geometrically very thick and maintains 
approximately uniform density to the boundary, so the wind is limited
to a small part of the fluid at the jet-disk interface. Therefore, for our 
purposes, we choose only to distinguish between the disk and funnel regions.

The simulation runs for a total time of $t_f  = 26548\, r_g / c$ and reaches
a quasi-steady state by time $t \approx 8000 r_g / c$. A snapshot of the fluid
data is saved every $\Delta t = 4 r_g / c$.
Modified spherical coordinates are used, with resolution 
$N_r \times N_\theta \times N_\phi = 272 \times 128 \times 256$.
This simulation is the highest resolution, longest duration 3D simulation of a 
MAD configuration to date.
The grid extends to a maximum radius of $R_\text{out} = 26000 r_g$.
In order to focus on the dynamics at small radii while avoiding 
numerical reflections off the outer boundary, the resolution is concentrated 
near the black hole, with a transition at $r = 500 r_g$ to a much sparser grid
\citep[see][for details]{MTB12}.
We limit our analysis to the inner $r = 200 r_g$, which corresponds to $194$ 
cells in the radial direction.
Coordinate singularities along the poles can cause further numerical 
difficulties and so we exclude cells near the poles from our radiative transport
calculations. This can be seen as an excised region along the $z$-axis in Figure 
\ref{fig:harm_data_thickdisk7}.

The jet in our MAD simulation is highly collimated by pressure support from
the geometrically very thick disk, and remains nearly cylindrical out to 
the boundary at $r = 200 r_g$.
For comparison, we checked our results against the 
A0.99N100 model from \citet{MTB12}.
This model is a MAD RIAF and is qualitatively similar to the fiducial model, 
however, the disk is geometrically thinner. We find similar spectra in both 
cases, indicating that our results are not just peculiarities of the very thick 
disk.

\begin{figure}
    \includegraphics[width=\harmfigfactor\linewidth]{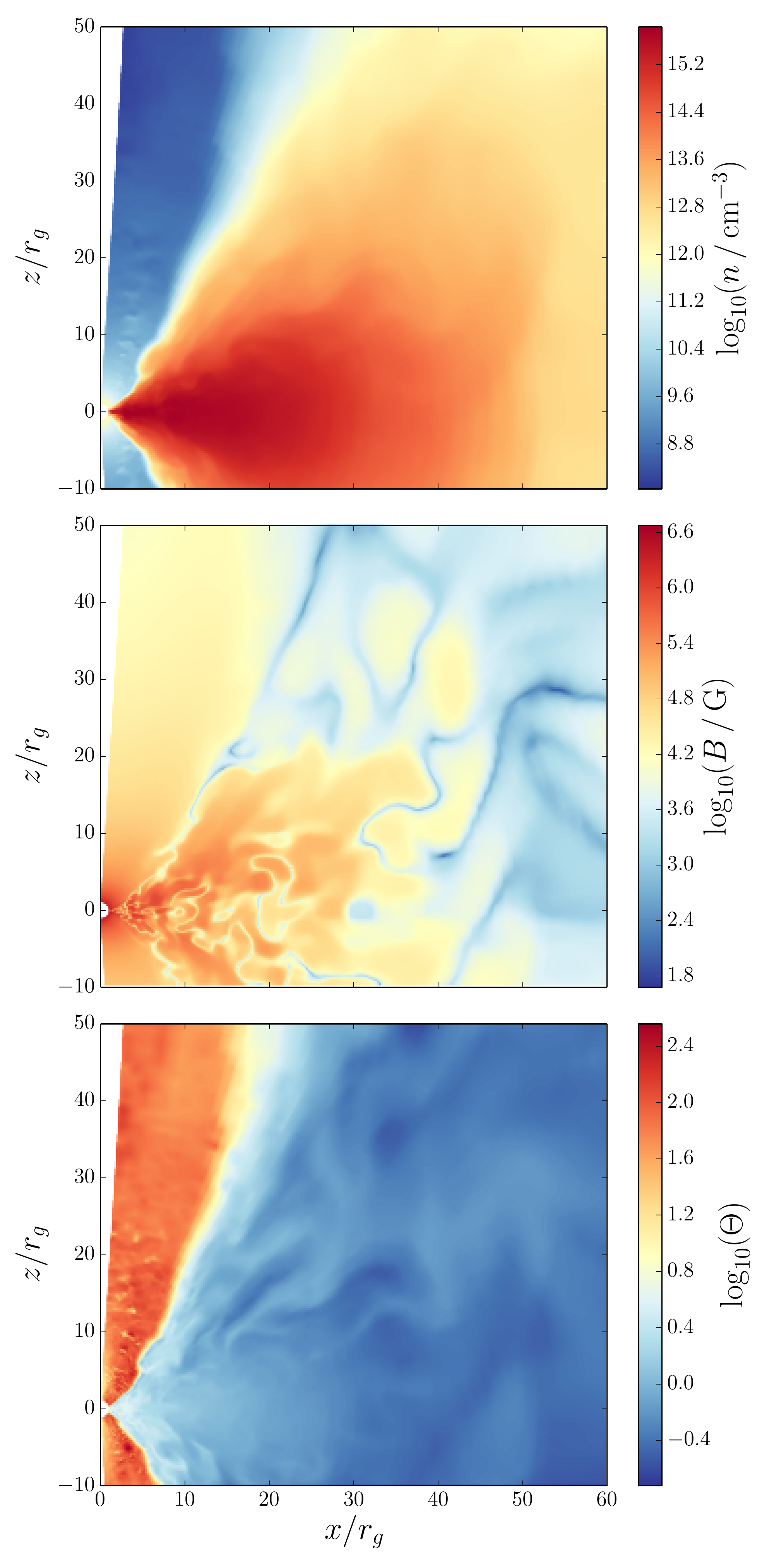}
    \caption{Electron number density, magnetic field strength, and 
        electron temperature, close to the
        black hole, at $t = 4000 r_g/c$, in our non-MAD model. 
        The inner disk is geometrically thicker and cooler than in the MAD 
        case.} 
    \label{fig:harm_data_nonMAD}
\end{figure}

For our non-MAD model, we use the dipole model of \citet{McB09}.
In this simulation, a MAD state does not develop and the accretion is driven
by the magneto-rotational instability.
In Figure \ref{fig:harm_data_nonMAD} we show snapshots of the electron number
density, magnetic field, and electron temperature at $t = 4000 r_g/c$, using 
the same parameters as in Figure \ref{fig:harm_data_thickdisk7}.
The black-hole magnetosphere does not disrupt the inner accretion 
flow in this case, and so the inner disk is geometrically thicker 
than in the MAD simulation. 
While the jet efficiency in our MAD simulation is $> 100$ per cent, the
corresponding efficiency in our non-MAD simulation is only about $1$ per cent, 
even with a large black-hole spin of $a = 0.92$.

The initial disk torus has inner edge at $r = 6 r_g$, pressure maximum 
at $r = 12 r_g$, and contains a single magnetic field loop.
The simulation runs for a total time of $t_f = 5000 r_g/c$ and reaches a 
quasi-steady state by time $t \approx 3000 r_g/c$.
The grid resolution is 
$N_r \times N_\theta \times N_\phi = 256 \times 128 \times 32$, and warps to
follow the disk at small radii and the jet at large radii.
The outer boundary is located at $R_\text{out} = 1000 r_g$. Again, we 
limit our calculations to the inner $r = 200 r_g$ and excise cells near the 
poles. We distinguish between the jet and disk using the same condition on
$b^2 / \rho c^2$ as in the MAD case.

\subsubsection{Density floors}
\label{subsubsec:floors}
The \texttt{HARM} code, as well as many other GRMHD codes 
(e.g., \texttt{WhiskyMHD}, \citeauthor{GR07} \citeyear{GR07};
\texttt{HARM3D}, \citeauthor{Noble+09} \citeyear{Noble+09};
\texttt{KORAL}, \citeauthor{Sadowski+13} \citeyear{Sadowski+13}; 
\texttt{IllinoisGRMHD}, \citeauthor{Etienne+15} \citeyear{Etienne+15};
\texttt{Athena++}, \citeauthor{WS15} \citeyear{WS15}),
can not handle a vacuum. If the rest-mass density
$\rho$, or the internal energy density $u$ become too small in comparison 
with $b^2$, truncation errors in the evolution can lead to large fractional
errors in these quantities. 
To avoid this, GRMHD codes use density ``floors'', which 
effectively inject mass into the system in regions where these floors are 
activated. 

In the simulations considered here,
the internal energy is chosen to enforce $u/\rho c^2 \leq 50$, then $\rho$ 
is chosen with the conditions that $b^2 / \rho c^2 \leq 50$ and 
$b^2 / u \leq 10^3$. 
We find numerically that these floors are only activated in the central 
regions of the highly-magnetized, 
low-density funnel. \citet{MG04} showed that, as long as 
$b^2 / \rho c^2 \gg 1$, the flow is approximately force-free 
(with maximum deviations of $\sim\text{few}\,\%$)
and so the dynamics of the electrodynamic field in 
the funnel is unaffected by the injection.    

Artificial mass injection primarily occurs near $r \sim 10 r_g$. 
At larger radii, this mass injection no longer occurs and the solution 
becomes a valid MHD solution, as shown in \citet{McKinney06}. 
The only effect of the
floors on the dynamics is therefore to set a rough upper limit on the bulk
Lorentz factor of $\Gamma_\text{max} = b^2/\rho c^2$ at large radii. 
In this work, we limit our analysis to the inner $r =200\,r_g$, where the 
Lorentz factor of the flow is much less
than the local value of $b^2/\rho c^2$. Therefore, the values
chosen for the floors do not have any effect on the dynamics of the 
jet in the simulated region.

Although the artificially injected material has no effect on the 
dynamics, it is potentially very hot and so could modify the
predicted spectra by overproducing high-energy emission.
Physically-motivated estimates of mass injection in funnel region    
suggest that the electron number density is in fact very low 
\citep{Moscibrodzka+11,LR11} and so should not contribute significantly to 
the emission \citep{MF13,Moscibrodzka+14}. 

To ensure that the injected mass does not affect the resulting spectra, we
remove this material before performing the radiative transport calculation 
on our non-MAD model.
For our MAD model, we found no need to remove this material, since
emission from regions which are potentially affected by the floors 
($b^2/\rho c^2 \gtrsim 10$) is negligible in this case 
(see Section \ref{subsubsec:jet sig MAD} for details).

\subsection{Radiative transport}
\label{subsec:rad transport}
We calculate the spectra and variability properties of the 
low/hard state in XRBs using a general relativistic radiative transport code 
based on the freely available \texttt{grmonty} \citep{Dolence+09}.
This code uses a post-processing approach for calculating the spectra and
relies on an external fluid model to supply the rest-mass density $\rho$,
internal energy density $u$, fluid four-velocity $u^\mu$, and magnetic field 
four-vector $b^\mu$, at every point in the grid. We interpolate these quantities
to arbitrary points as needed.
We modify the original code to work with general 3D \texttt{HARM} data as 
input, and to allow for different temperature prescriptions in the disk and in 
the jet (see Section \ref{subsubsec:electron temperature}). 

The spectra are calculated assuming synchrotron emission, self-absorption, and 
Compton scattering from a thermal distribution of relativistic electrons. 
The distribution function for relativistic electrons at temperature $\Theta$ is
\begin{equation}
    \label{eq:distribution}
    \frac{\diff n}{\diff \gamma} = \frac{n}{\Theta}
    \frac{\gamma^2 \beta}{K_2(\Theta^{-1})}
    \exp\left(-\frac{\gamma}{\Theta}\right)
\end{equation}
where $n$ is the number density of electrons,
$\gamma = (1 - \beta^2)^{-1/2}$ is the electron Lorentz factor, 
$\beta$ is the electron speed in the fluid frame, and $K_2$ is the modified
Bessel function of the second kind. We neglect any radiative cooling of the
electrons and so the electron distribution function is determined, at every 
point in the grid, by the local fluid properties.
We use the following emissivity for thermal synchrotron emission, valid for
$\Theta \gtrsim 0.5$ \citep[see][]{Dolence+09}
\begin{subequations}
    \label{eq:emissivity}
    \begin{equation}
        j_\nu = \frac{\sqrt{2}\pi e^2 n \nu_s}{3cK_2(\Theta^{-1})}
        \left(X^{1/2} + 2^{11/12} X^{1/6} \right)^2 \exp(-X^{1/3})    
    \end{equation}    
    \begin{equation}
        X \equiv \frac{\nu}{\nu_s}
    \end{equation}    
    \begin{equation}
        \nu_s \equiv \frac{2}{9} \left(\frac{eB}{2\pi m c} \right) \Theta^2
        \sin\theta
    \end{equation}    
\end{subequations}
where $e$ is the electron charge, $B$ is the magnetic field strength, 
and $\theta$ is the angle between the photon wave vector and the magnetic field.
The absorption coefficient is calculated as
\begin{equation}
    \label{eq:absorption}
    \alpha_{\nu,a} = \frac{j_\nu}{B_\nu}
\end{equation}
where $B_\nu$ is the Planck function.
The extinction coefficient for Compton scattering from a distribution of
relativistic electrons is given by
\begin{equation}
    \label{eq:extinction}
    \alpha_{\nu,s} = n \sigma_h
\end{equation}
where $\sigma_h$ is the ``hot cross section'' defined as
\begin{equation}
    \label{eq:hotcross}
    \sigma_h \equiv \frac{1}{n} \int \diff^3 p \frac{\diff n}{\diff^3 p}
    (1 - \mu \beta) \sigma_\text{KN}
\end{equation}
Here, $p$ is the electron four-momentum, 
$\diff^3 p = \diff p_1 \diff p_2 \diff p_3$, and $\mu$ is the cosine of the 
angle between the electron momentum and photon momentum in the fluid frame. The
Klein-Nishina cross section, $\sigma_\text{KN}$, is
\begin{equation}
    \label{eq:KN}
    \sigma_\text{KN} = \sigma_T \frac{3}{4\epsilon^2}
    \left(2 + \frac{\epsilon^2 (1 + \epsilon)}{(1+2\epsilon)^2}
        +\frac{\epsilon^2 - 2\epsilon - 2}{2\epsilon} 
        \log(1 + 2\epsilon)\right)
\end{equation}
where $\sigma_T$ is the Thomson cross section, and 
$\epsilon = \epsilon' \gamma (1 - \mu \beta)$ is the photon energy 
(in units of $mc^2$) in the electron rest frame, and $\epsilon'$ is the photon 
energy in the fluid frame. 
We use the thermal distribution in equation \eqref{eq:distribution} when 
calculating the hot cross section \eqref{eq:hotcross}.
The scattering calculation samples the Klein-Nishina differential cross section
\begin{equation}
    \label{eq:diffKN}
    \frac{2\pi}{\sigma_T} \frac{\diff\sigma_\text{KN}}{\diff \epsilon_s}
    = \frac{1}{\epsilon_s} \left(\frac{\epsilon}{\epsilon_s} + 
    \frac{\epsilon_s}{\epsilon} - 1 + \cos^2\theta_s \right)
\end{equation}
where $\epsilon_s$ is the energy of the scattered photon, and $\theta_s$ is the 
scattering angle in the electron frame.

Introducing radiation breaks the scale-free nature of the GRMHD data.
We set the length and time scales by specifying the black hole mass $M$.
The appropriate scales are then the
gravitational radius, $r_g$, and the light crossing time, $t_g = r_g / c$.
The fluid mass/energy unit $\mathcal{M}$ must also be specified
(this is not set by $M$ because the fluid mass is $\ll M$). 
Using these units, the \texttt{HARM} data can be scaled to a particular system, 
for example, the mass density is set as $\rho = (\mathcal{M} / r_g^3)
\widetilde{\rho}$, where $\widetilde{\rho}$ is the dimensionless mass density 
given by the \texttt{HARM} code. 
Note that once $M$ is chosen, the accretion rate at a given radius is set by
$\mathcal{M}$ via
\begin{equation}
    \dot{M} = \left| \int \sqrt{-g} \diff x^\theta \diff x^\phi \rho u^r \right|
\end{equation}
For our purposes, we set $M = 10M_\odot$ and
choose $\mathcal{M}$ such that the accretion rate at the black-hole
horizon is $\dot M = 10^{-5} \dot{M}_\text{Edd}$.

By tracking photons individually, we can unambiguously 
determine the jet contribution to the spectrum. 
We track $\sim 10^8$ photons to an outer radial boundary of $r=200r_g$. 
The choice of this boundary is discussed in Section \ref{subsec:grmhd sim} and 
has little effect on the results as most of the high-energy emission originates 
close to the black hole. 
While relativistic Doppler effects are fully accounted for by the code, 
we find that the effects on the resulting spectra are 
small since the jets in our simulations are only mildly relativistic at small
radii.

For computational simplicity, we use a ``fast light'' approximation in which 
the fluid data is treated as time-independent during the radiative transport 
calculation. This approximation may break down in regions where the 
light crossing time is comparable to the dynamical time, however, we perform our 
post-processing calculation only after the fluid simulation has
reached a quasi-steady state and so we expect this to be a reasonable 
approach.
Furthermore, \citet{Shcherbakov+12} performed both time-independent and fully
time-dependent radiative transport calculations in the context of Sgr A*, and
found good agreement in most cases.

\subsubsection{Disk and jet electron temperatures}
\label{subsubsec:electron temperature}
The details of the electron thermodynamics in RIAFs have not been determined.
A common approach is to assume that the electron temperature is some constant
fraction of the proton temperature, and to use this ratio as a free parameter
\citep{Moscibrodzka+09}.
Although more sophisticated models are being developed 
\citep{Ressler+15,Foucart+15}, 
there are still many parameters whose values are unknown.
Because of these uncertainties, we use the simple assumption of a constant 
proton-to-electron temperature ratio $T_p/T_e$.
However, since differences in density and magnetization in the disk and 
jet can lead to different cooling rates for the electrons in these regions, 
we vary this temperature ratio independently in these regions 
\citep{Chan+15a,Ressler+15}. 
We define a proton-to-electron temperature ratio $\R_d$ in the disk where
$b^2/\rho c^2 < 0.5$, and a ratio $\R_j$ in the jet where 
$b^2/\rho c^2 \geq 0.5$.

The values of these ratios depend on poorly understood electron thermodynamics. 
However, assuming that (i) the dissipation of turbulence mainly heats the 
protons, (ii) the cooling time for the electrons is shorter than that of the 
protons, and (iii) the electron cooling time is shorter than the timescale for 
significant energy exchange between the electrons and protons, we expect these 
temperature ratios to be greater than unity \citep{YN14,Chan+15a}.
Furthermore, because of the similarities between AGN and the low/hard state in 
XRBs, we assume that the physics of electron heating and cooling is the same 
across these systems.
We therefore choose a range of values of $\R_d$ and $\R_j$ motivated by fitting
to Sgr A* and M87, since these are the only sources whose spectra have been 
fitted to constrain these parameters
\citep[][]{Moscibrodzka+09,MF13,Moscibrodzka+14,Chan+15a,Moscibrodzka+15}.

\vfill

\section{Results}
\label{sec:results}
\subsection{Jet signatures}
\label{subsec:jet sig}
\subsubsection{MAD model}
\label{subsubsec:jet sig MAD}
\begin{table}
    \centering
    \begin{tabular} {c |c c}
        \hline \hline
        Model & $\R_d$ & $\R_j$ \\ \hline
        1 & 3 & 3 \\
        2 & 10 & 10 \\
        3 & 30 & 30 \\
        4 & 3 & 10 \\
        5 & 3 & 30 \\
        6 & 10 & 30 \\
        7 & 10 & 3 \\
        8 & 30 & 3 \\
        9 & 30 & 10 \\
        \hline
    \end{tabular}
    \caption{List of MAD model proton-to-electron temperature ratios.}
    \label{table:MAD models}
\end{table}

\begin{figure}
    \includegraphics[width=\figfactor\linewidth]{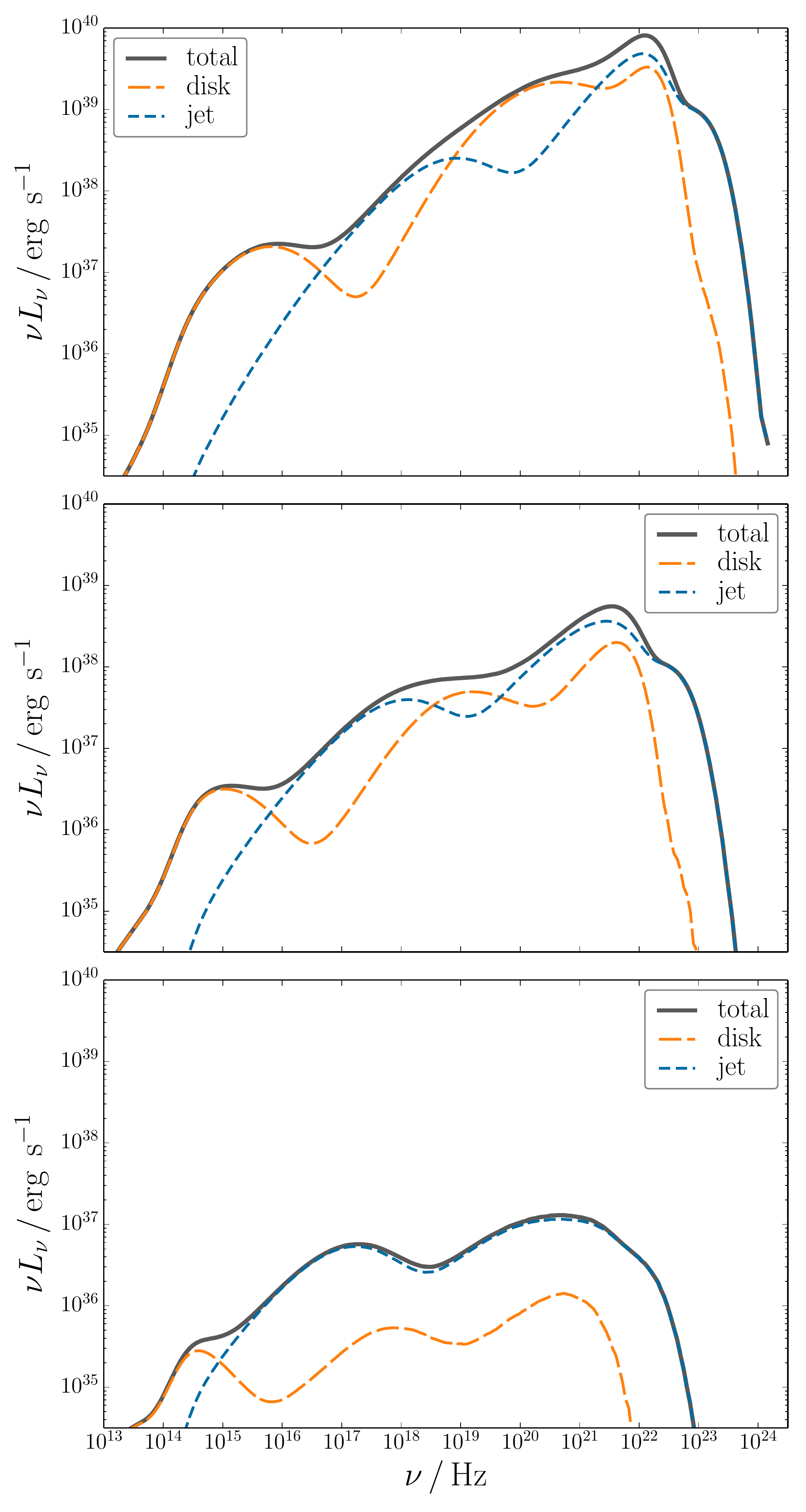}
    \caption{MAD model spectra with $\R_d = \R_j$. 
        From top to bottom, these were calculated with 
        $(\R_d, \R_j) = (3, 3), (10, 10), (30, 30)$, 
        respectively. The disk contribution dominates mainly around 
        $10^{15}\,\text{Hz}$, while the jet contributes significantly in the 
        X-rays and $\gamma$-rays.}
    \label{fig:Rd_eq_Rj}
\end{figure}

For our MAD model, we calculate spectra for the nine temperature models listed 
in Table \ref{table:MAD models}. 
In Figure \ref{fig:Rd_eq_Rj} we show the spectra calculated with 
$\R_d = \R_j$.
The distinction between the jet and disk contributions is 
defined such that the ``jet'' (short dashes) component corresponds
to the contribution from photons which either originated in the jet or
scattered in the jet before escaping. The ``disk'' (long dashes)
component corresponds to photons which originated in the disk and escaped 
without scattering in the jet (possibly scattering in the disk before leaving  
the system).

The middle panel shows the spectrum calculated with 
$(\R_d, \R_j) = (10,10)$. 
This spectrum qualitatively captures the main spectral
features present in most models, which we describe below.
Both the ``disk'' and ``jet'' components have three peaks.
The peak in the ``disk'' component at $\sim 10^{15}\,\text{Hz}$ is due to
synchrotron emission from the disk, while the two higher peaks 
at $\sim 10^{19}\,\text{Hz}$ and $\sim 10^{22}\,\text{Hz}$ result from 
single and double synchrotron self-Compton, respectively.
The peak in the ``jet'' component at $\sim 10^{18}\,\text{Hz}$ is
due to synchrotron emission from the jet, while the peak at 
$\sim 10^{22}\,\text{Hz}$ corresponds to synchrotron photons from the jet which 
scattered once in the disk before escaping. The peak at 
$\sim 10^{23}\,\text{Hz}$ is due to single scattering in the jet.
In all models with $\R_d = \R_j$, 
the disk dominates in the optical, while the jet contributes significantly to 
the X-rays and $\gamma$-rays. 
The disk contributes to the hard X-rays in models with $\R_d < 30$.
In these models, the disk emission peaks around $10^{22} \,\text{Hz}$, 
and decays rapidly above this.
The emission decays since the photons have
been scattered up to the same temperature as the electrons in the disk.
In what follows, we will refer to this frequency as the 
``saturation frequency'', $\nu_\text{sat}$.

It is interesting to note that, although all these models have $\R_d = \R_j$, 
there are differences in the resulting spectra.
This is due to the strong dependence of the scattering on the electron 
temperature. The synchrotron peak depends on
the temperature as $(\nu j_\nu)_\text{syn} \sim \Theta^2$, while the inverse
Compton peak goes like $(\nu j_\nu)_\text{IC} \sim y(\nu j_\nu)_\text{syn} 
\sim \Theta^4$. Here, $y$ is the Compton $y$ parameter given by 
$y = 16 \Theta^2 \tau$ \citep{RL79}, and $\tau$ is the optical depth.
We have assumed that the fluid is optically thin, and that the 
electrons are ultrarelativistic, $\gamma \gg 1$, and have a thermal 
distribution.

\begin{figure}
    \includegraphics[width=\figfactor\linewidth]{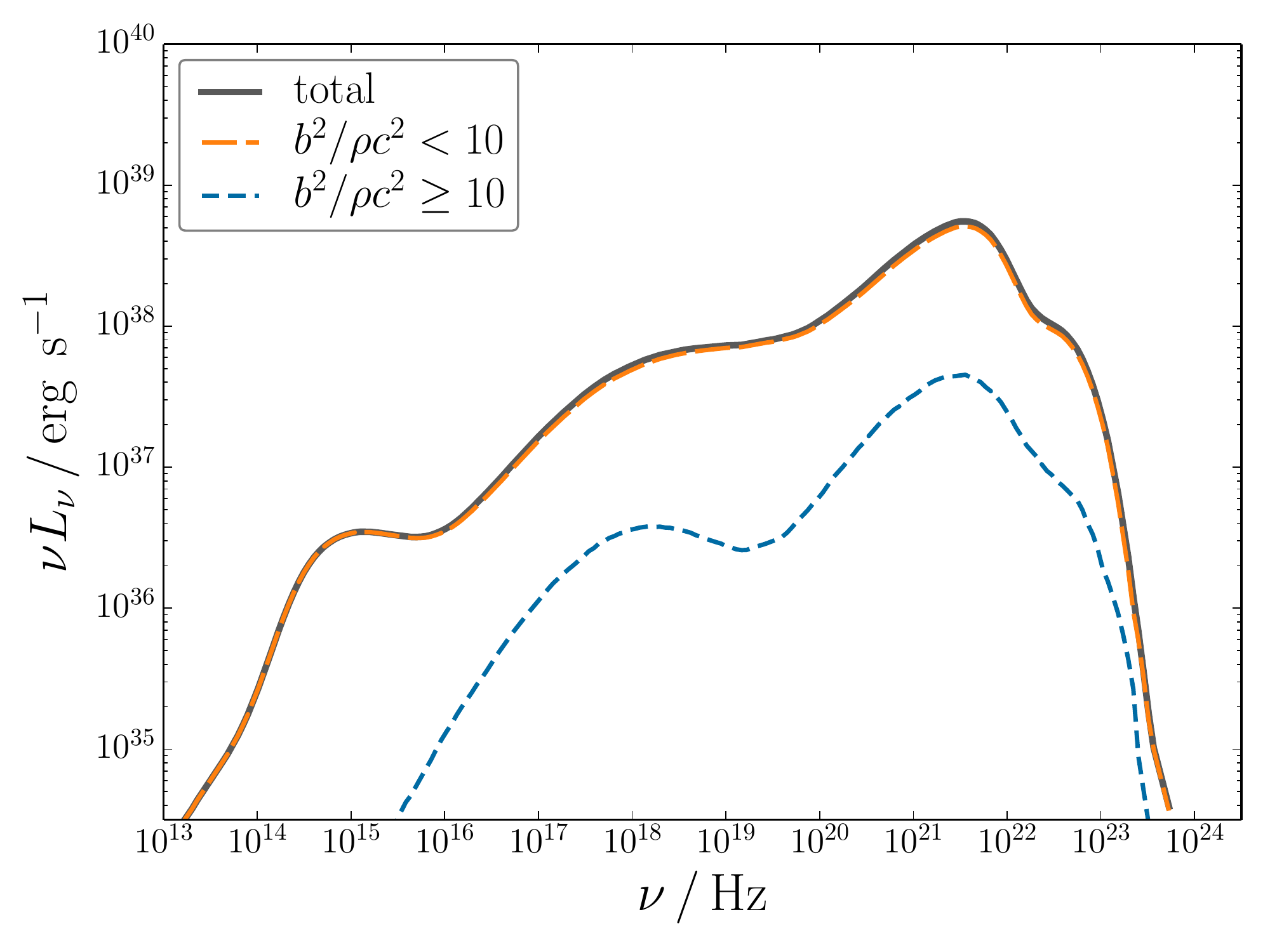}
    \caption{MAD model spectrum with $(\R_d, \R_j) = (10, 10)$.
        It is clear that emission from regions with $b^2/\rho c^2 \geq 10$ has
        little effect on the overall spectrum.}
    \label{fig:MAD_floors}
\end{figure}

In Figure \ref{fig:MAD_floors} we show the effects of the floors on the spectrum 
calculated with $\R_d = \R_j = 10$ (middle panel of Figure \ref{fig:Rd_eq_Rj}).
Although mass is initially injected with $b^2/\rho c^2 = 50$, 
at late times any region with $b^2/\rho \gtrsim 10$ is 
likely dominated by floor material.
It is clear from Figure \ref{fig:MAD_floors} that this injected mass
($b^2/\rho c^2 \geq 10$)
is $\sim 1.5$ orders of magnitude less luminous than the rest of the plasma
($b^2/\rho c^2 < 10$), and so we conclude that the floors
have little effect on the predicted spectra from our MAD model.

\begin{figure}
    \includegraphics[width=\figfactor\linewidth]{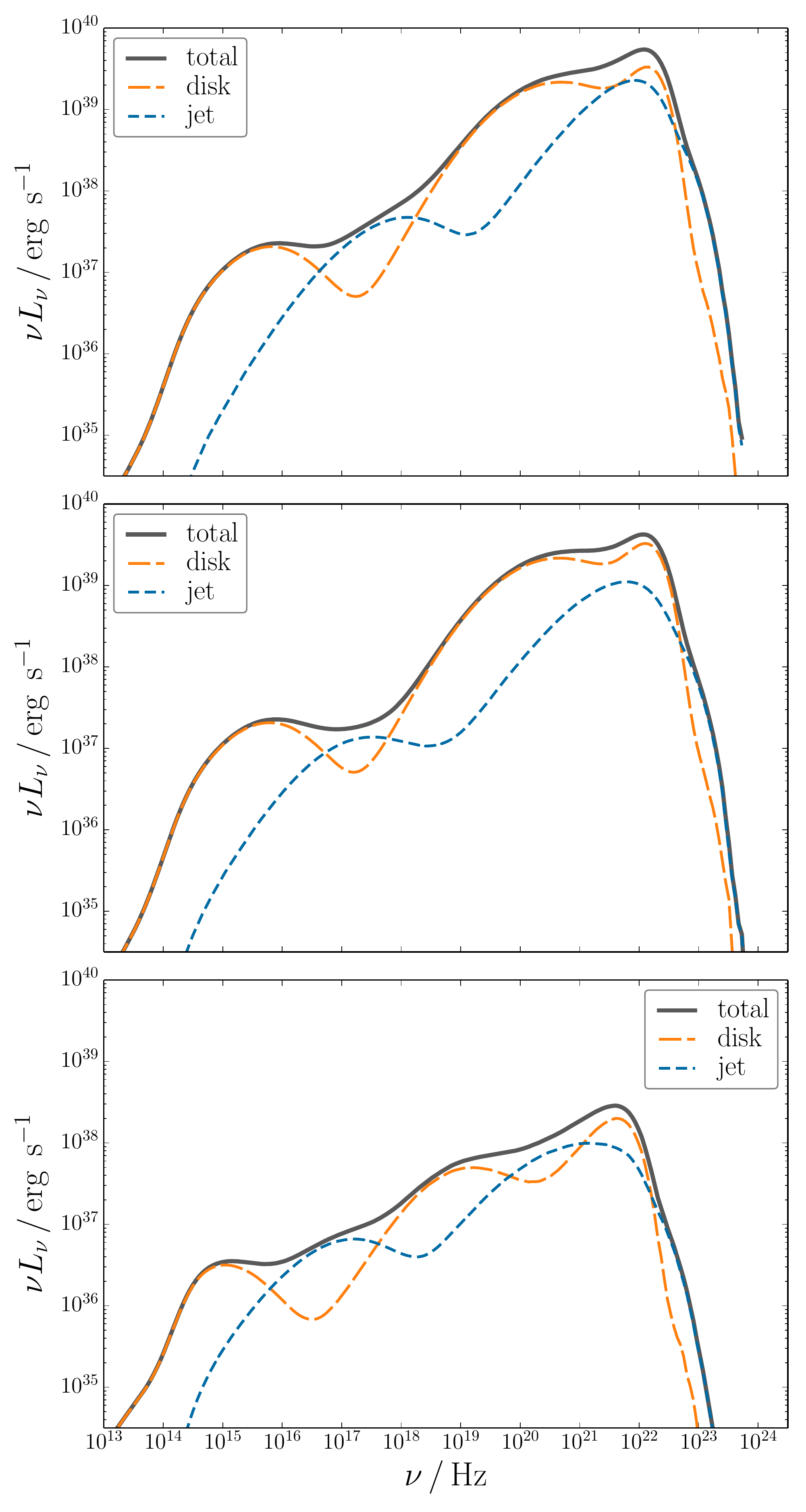}
    \caption{Same as Figure \ref{fig:Rd_eq_Rj} but for models with 
        $\R_d < \R_j$. 
        From top to bottom 
        $(\R_d, \R_j) = (3, 10), (3, 30), (10, 30)$,
        respectively. The disk dominates the spectra at all wavelengths in this 
        case. However,there are frequencies where the jet contributes 
        significantly.}
    \label{fig:Rd_less_Rj}
\end{figure}

In Figure \ref{fig:Rd_less_Rj} we show spectra calculated with 
$\R_d < \R_j$. 
The features in the ``disk'' component are similar to those in Figure 
\ref{fig:Rd_eq_Rj}, with a synchrotron peak around $\sim 10^{15}\,\text{Hz}$, 
and two higher energy peaks due to single and double synchrotron self-Compton.
The ``jet'' component shows a synchrotron peak at
$\sim 10^{18}\,\text{Hz}$, and a peak at $10^{22}\,\text{Hz}$ corresponding 
to photons which originated in the jet and scattered once in the disk before 
escaping. The disk dominates most of the spectra in this case. 
The high-energy $\gamma$-ray peak, present in models with $\R_d = \R_j$, is 
absent or obscured by the hotter disk contribution.

\begin{figure}
    \includegraphics[width=\figfactor\linewidth]{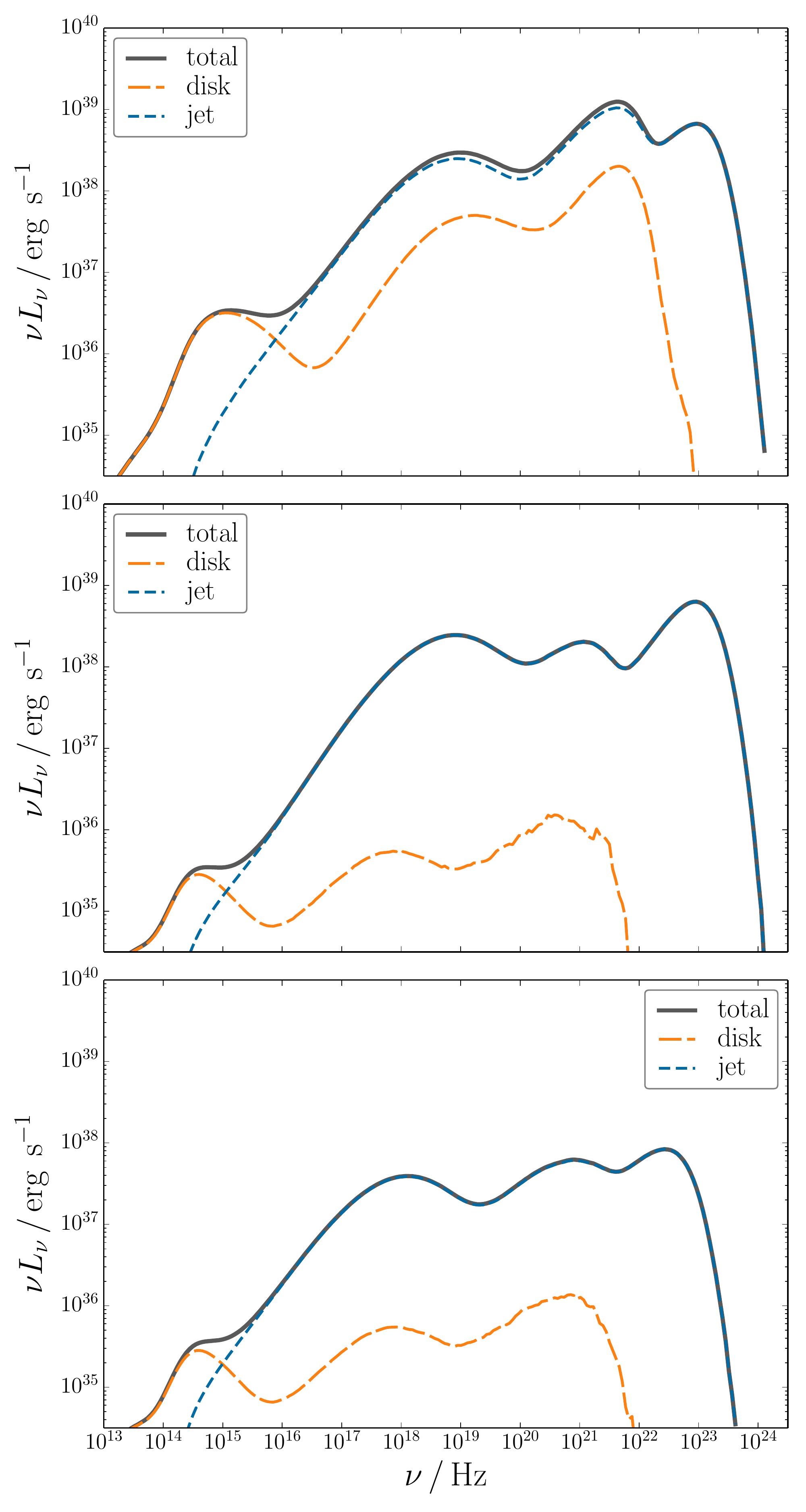}
    \caption{Same as Figure \ref{fig:Rd_eq_Rj} but for models with 
        $\R_d > \R_j$. 
        From top to bottom
        $(\R_d, \R_j) = (10, 3), (30, 3), (30, 10)$,
        respectively. The jet dominates everything above 
        $\sim 10^{16}\,\text{Hz}$.}
    \label{fig:Rd_gtr_Rj}
\end{figure}

In Figure \ref{fig:Rd_gtr_Rj} we show spectra calculated with 
$\R_d > \R_j$. 
In this case, the jet dominates most of the
spectrum, with a small contribution from the disk around the optical band.
The peak around $\sim 10^{15}\,\text{Hz}$ is due to synchrotron from the disk,
while the peak at $\sim 10^{19}\,\text{Hz}$ is synchrotron emission from the 
jet. The third peak, at $\sim 10^{21}\,\text{Hz}$, again corresponds to photons 
which were emitted in the jet and scattered once in the disk.
The peak in the $\gamma$-rays around $10^{23}\,\text{Hz}$ is due to scattering 
in the jet.

The locations of the synchrotron and saturation peaks provide a wealth of 
information about the fluid properties in the jet and in the disk.
The ratio of the jet and disk synchrotron peak frequencies depends on the 
temperatures and magnetic fields as
$\nu_{\text{syn},j}/\nu_{\text{syn},d} \sim \Theta_j^2 B_j/\Theta_d^2 B_d$.
The saturation frequency is simply proportional to the electron temperature, 
$\nu_\text{sat} \sim \Theta$.
Therefore, the ratio of jet and disk magnetic fields can be estimated from
the spectra as
\begin{equation}
    \label{eq:B_ratio}
    \frac{B_j}{B_d} \sim 
    \left(\frac{\nu_{\text{syn},j}}{\nu_{\text{syn},d}}\right)
    \left(\frac{\nu_{\text{sat},d}}{\nu_{\text{sat},j}}\right)^2
\end{equation}
For example, the top panel of Figure \ref{fig:Rd_gtr_Rj} shows 
$\nu_{\text{sat},d}/\nu_{\text{sat},j} \sim 1/30$ and 
$\nu_{\text{syn},j}/\nu_{\text{syn},d} \sim 10^4$, which corresponds to a
magnetic field ratio of $B_j/B_d \sim 10$. This analysis is independent of the
temperature model, however, we have used the fact that the jet in our simulation 
is only mildly relativistic.

While separating the spectrum into jet and disk components is useful for
identifying their contributions, in reality, this decomposition is not so
straightforward. Therefore, we are interested in identifying signatures 
of jet emission in the composite spectrum. 

In all our MAD calculations, the highest energy emission is produced by inverse 
Compton scattering of synchrotron photons. Therefore, the electron temperature 
sets an upper limit on the high energy emission. In all models with 
$\R_d \geq \R_j$ 
(Figures \ref{fig:Rd_eq_Rj} and \ref{fig:Rd_gtr_Rj}) the jet electrons are one 
or two orders of magnitude hotter than those in the disk. 
Therefore, we expect the highest energy emission to come from the jet.
This is clearly visible in the
spectra as a $\gamma$-ray peak in the jet component around 
$\sim 10^{23}\,\text{Hz}$, well above the highest energy disk contribution.
This feature is absent in disk-dominated spectra, i.e., those with
$\R_d < \R_j$ (see Figure \ref{fig:Rd_less_Rj}).
We conclude that this high-energy feature could be a good indicator of jet 
emission.

Another possible signature of jet emission occurs in regions where the 
spectra change from disk to jet dominated.
The overlapping jet and disk components tend to smooth out parts of the 
spectrum which would otherwise be much steeper. Most of the spectra from our MAD 
simulation show roughly flat ($\nu L_\nu \sim \nu^0$) regions, followed by a 
break where the spectrum changes to $\nu L_\nu \sim \nu$.
This can be seen clearly in the spectra in Figure \ref{fig:Rd_gtr_Rj},
with breaks around $\sim 10^{15}\,\text{Hz}$.
There is a second break in the spectrum around $\sim 10^{18}\,\text{Hz}$, where
it returns roughly to $\nu L_\nu \sim \nu^0$. 
This second break is followed by ``wiggles'' in spectrum, with variations in 
the luminosity of a factor of a few.
These features are less clear in models where the spectra are almost completely 
dominated by disk emission ($\R_d < \R_j$). 
The breaks are due to the combined effect of the jet and disk contributions,
and so are a clear indication of the presence of jet emission.

\subsubsection{Non-MAD model}
\label{subsubsec:jet sig nonMAD}
For our non-MAD model, we use the same black hole mass as in our MAD 
calculations.
Since we are interested in signatures of jets,
we choose temperature models which potentially show a substantial jet 
contribution, i.e., those with $\R_d > \R_j$.
For comparison with our MAD model, we choose  
$\dot{M} = 10^{-5} \dot{M}_\text{Edd}$.
In this case, the spectra are primarily dominated by disk emission and so
we also investigate a lower accretion rate of 
$\dot{M} = 10^{-6} \dot{M}_\text{Edd}$.

\begin{figure}
    \includegraphics[width=\nmfigfactor\linewidth]{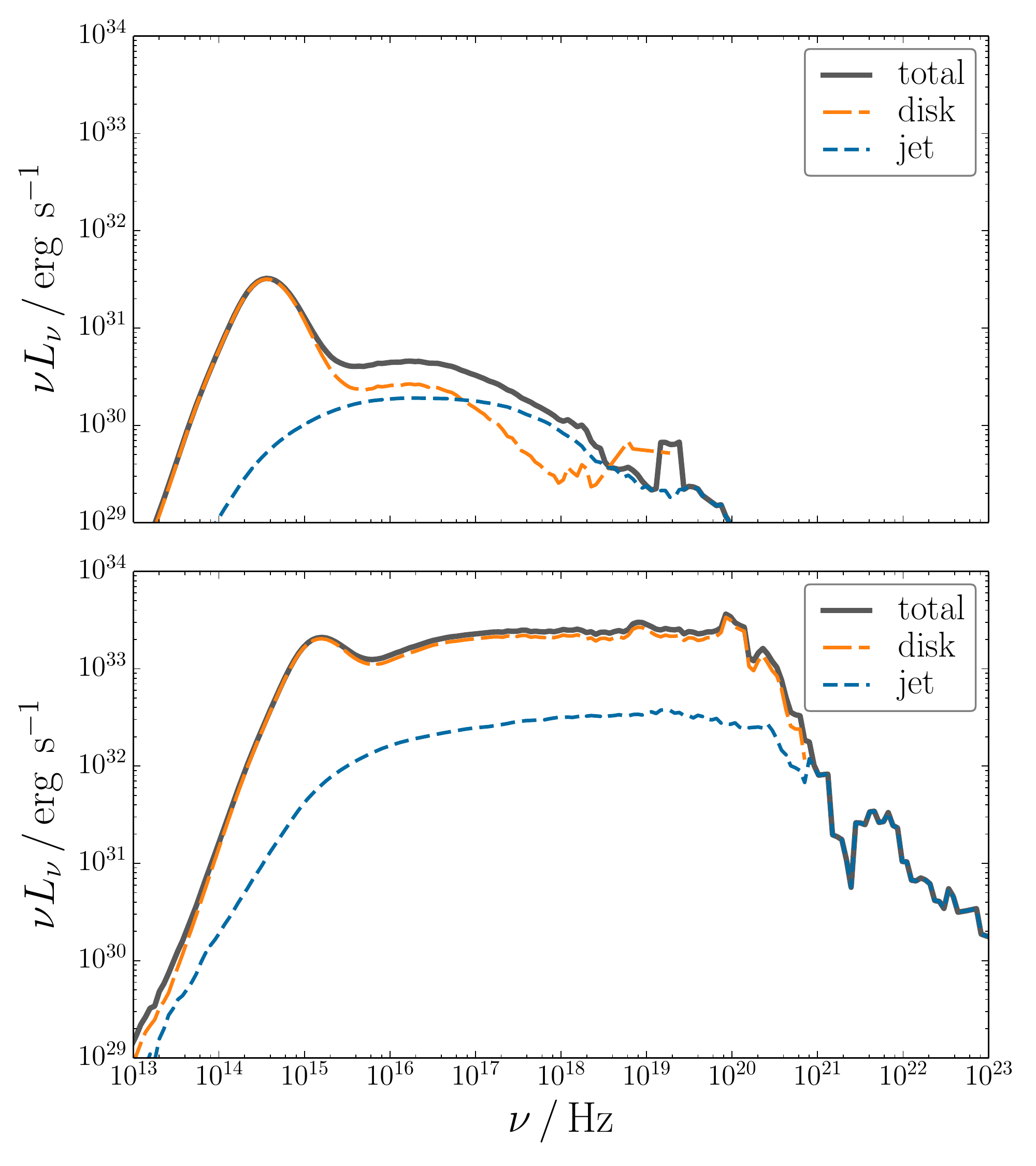}
    \caption{Non-MAD model spectra calculated with $(\R_d, \R_j) = (30, 3)$. 
        The top panel shows the spectrum with 
        $\dot{M} = 10^{-6}\dot{M}_\text{Edd}$, while the bottom panel 
        has $\dot{M} = 10^{-5}\dot{M}_\text{Edd}$. 
        High-energy $\gamma$-ray emission is clearly produced by the jet. 
        In models where the disk synchrotron peaks near or below 
        $\sim 10^{14}\,\text{Hz}$, the jet contributes significantly to the 
        X-rays. In models where this synchrotron peak is near or above 
        $\sim 10^{15}\,\text{Hz}$, the X-rays are dominated by emission from 
        the disk. 
        The higher energy emission is noisy due to poor photon statistics.}
    \label{fig:nonMAD_Mdot_30_3}
\end{figure}

In Figure \ref{fig:nonMAD_Mdot_30_3} we show spectra from our non-MAD model, 
calculated with $(\R_d, \R_j) = (30, 3)$ and accretion rates of 
$10^{-6} \dot{M}_\text{Edd}$ (top panel) and 
$10^{-5} \dot{M}_\text{Edd}$ (bottom panel).
These spectra show pronounced synchrotron peaks from the disk at 
$\sim 10^{14}\,\text{Hz}$ and $\sim 10^{15}\,\text{Hz}$. In the model with 
$\dot{M} = 10^{-6} \dot{M}_\text{Edd}$, the jet component contributes
significantly to the X-rays,
while in the model with $\dot{M} = 10^{-5} \dot{M}_\text{Edd}$, the 
disk dominates at all frequencies up to the $\gamma$-rays. 
Interestingly, although the disk component dominates most of the spectrum
in the $\dot{M} = 10^{-5} \dot{M}_\text{edd}$ case, 
there is significant $\gamma$-ray emission from the jet at and above
$\sim 10^{22}\,\text{Hz}$. 
As in our MAD model, this is due to scattering in the jet and is located at 
higher frequencies than the disk saturation frequency, i.e., 
above where the disk emission decays. 
From the top panel of Figure \ref{fig:nonMAD_Mdot_30_3}, we can conclude that a 
pronounced synchrotron peak at or below 
$\sim 10^{14}\,\text{Hz}$, which can be attributed to the disk, 
indicates that any observed X-ray emission is likely due to emission from the 
jet.

\begin{figure}
    \includegraphics[width=\nmfigfactor\linewidth]{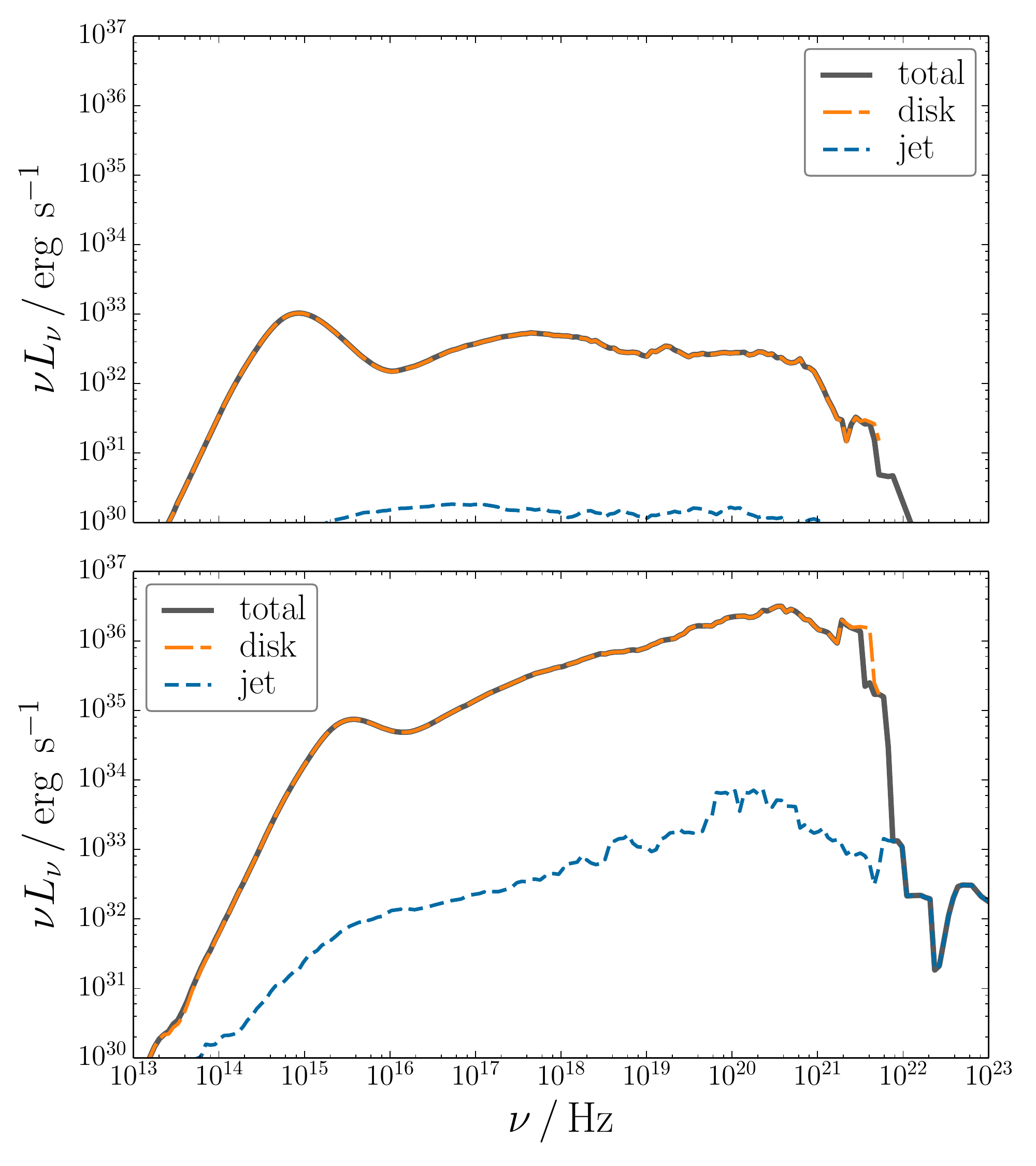}
    \caption{Non-MAD model spectra calculated with $(\R_d, \R_j) = (10, 3)$. 
        The top panel shows the spectrum with 
        $\dot{M} = 10^{-6}\dot{M}_\text{Edd}$, while the bottom panel 
        has $\dot{M} = 10^{-5}\dot{M}_\text{Edd}$. Synchrotron emission 
        and scattering from the disk dominate most of the spectrum. 
        Similar to the MAD case, 
        the high-energy $\gamma$-ray emission above $\sim 10^{22}\,\text{Hz}$ is 
        due to scattering in the jet.}
    \label{fig:nonMAD_Mdot_10_3}
\end{figure}

In Figure \ref{fig:nonMAD_Mdot_10_3} we show spectra calculated with the same
accretion rates as in Figure \ref{fig:nonMAD_Mdot_30_3}, but with 
$(\R_d, \R_j) = (10, 3)$. 
In this case, there is a peak at $\sim 10^{15}\,\text{Hz}$ due to synchrotron 
emission from the disk, while the rest of the spectrum up to 
$\sim 10^{21}\,\text{Hz}$ is dominated by synchrotron self-Compton from the 
disk. 
Again, the highest-energy $\gamma$-rays are produced by scattering in the jet.
Therefore, this is a robust signature of jet 
emission which is independent of whether the accretion flow is MAD or non-MAD.
It is interesting to note that the X-rays from our MAD model are 
dominated by synchrotron photons from the jet, while the
X-rays are produced by scattering in the disk in our non-MAD model 
(see Figures \ref{fig:Rd_gtr_Rj} and \ref{fig:nonMAD_Mdot_10_3}).

\subsection{MAD model variability}
\label{subsec:variability}
In this Section, we investigate jet variability in our MAD model, 
and so choose a temperature model which produces significant jet emission.
In what follows we set $\R_d = 10$, and $\R_j = 3$.
\subsubsection{Magnetic field inversion}
\label{subsubsec:magnetic field inversion}
The initial magnetic field in our MAD model contains multiple 
poloidal field loops, with adjacent field loops having opposite polarity.
\citet{Igumenshchev09} argued that the accretion of such oppositely polarized 
loops could be responsible for the observed state transitions in XRBs.
As discussed in \citet{Dexter+14}, the polarity inversion causes 
large-scale magnetic reconnection in the disk.
The inner disk, compressed by the black-hole magnetosphere 
in the MAD state, expands vertically due to the decreasing magnetic pressure.
During the inversion (a timescale of $\sim 2000 r_g/c \sim 0.1$s), the MAD state
is destroyed and the disk more closely resembles that of our non-MAD model, in
which the accretion is driven by the magneto-rotational instability. 
The steady BZ jet is also quenched by this process and a new
transient jet is launched by the reconnecting field.
This transient jet is mildly relativistic, with velocity $\sim 0.1c$ at 
$200 r_g$, and is qualitatively similar to the transient, ballistic jets seen
during transitions from the hard to soft state.

\begin{figure}
    \includegraphics[width=\linewidth]{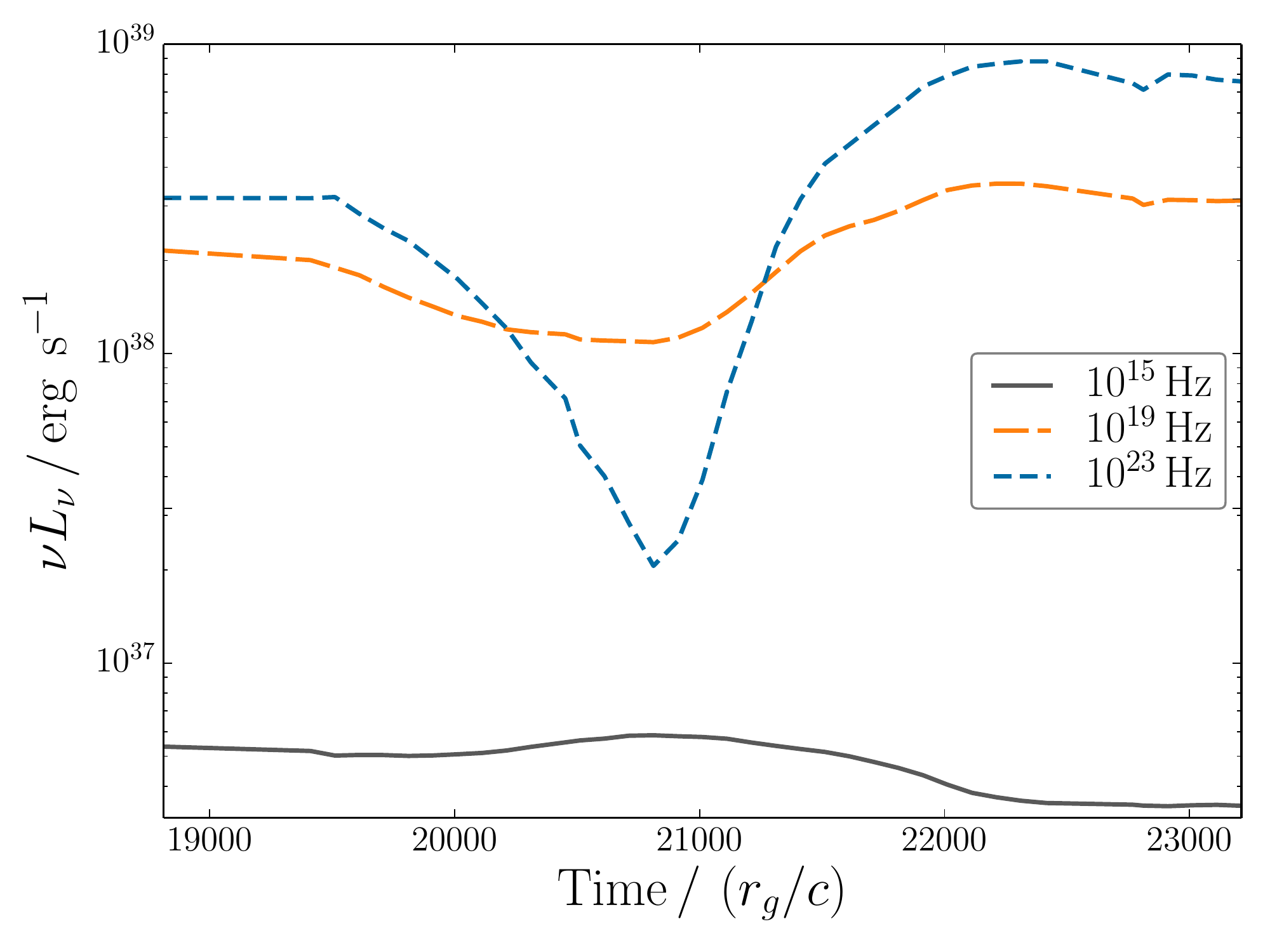}
    \caption{Optical, X-ray, and $\gamma$-ray variability during a global
        magnetic field inversion.  These lightcurves were calculated with
        $\R_d = 10,\, \R_j = 3$. The $\gamma$-ray luminosity
        varies by nearly two orders of magnitude, and the ratio of the
        $\gamma$-ray to X-ray luminosities, $L_\gamma / L_X$, varies such that 
        $L_\gamma / L_X > 1$ in the steady BZ jet and $L_\gamma / L_X < 1$ 
        during the transient outburst.}
    \label{fig:inversion}
\end{figure}

Here, we investigate the observational signatures of such a polarity inversion.
In Figure \ref{fig:inversion}, we show the evolution of the optical, X-rays,
and $\gamma$-rays during the global magnetic field inversion in which the
MAD state is destroyed and then re-established.
In the initial MAD state ($t\approx19000\, r_g/c$), the optical band is 
dominated by synchrotron emission from the disk, while the X-rays and $\gamma$ 
rays are produced by synchrotron emission and Compton scattering in the steady 
BZ jet. In this state, the ratio of the $\gamma$-ray to X-ray luminosities is 
$L_\gamma / L_X > 1$. 
During the transient outburst, corresponding to the destruction of the MAD state, 
this ratio changes to $L_\gamma / L_X < 1$.  After the inversion, the disk 
returns to a MAD state and  the BZ jet is re-launched with $L_\gamma / L_X > 1$. 

Overall, the $\gamma$-ray luminosity varies by
nearly two orders of magnitude while the X-rays vary by a factor of a few.
There is a small increase in optical emission from the disk, peaking around the
minimum of the $\gamma$-ray and X-ray emission. The re-launched BZ
jet is significantly more luminous in the $\gamma$-rays and X-rays, while the
disk is less luminous after the outburst.
The X-ray and $\gamma$-ray lightcurves, and in particular the ratio 
$L_\gamma / L_X$, could be used as an observational probe of such a global 
magnetic field inversion, and so might be useful for directly comparing models 
of state transitions in XRBs with observations.

\subsubsection{Jet-disk quasi-periodic oscillations}
\label{subsubsec:jet-disk qpos}
\citet{MTB12} found that the black hole magnetosphere and disk exhibit 
significant quasi-periodic oscillations (QPOs) in dynamical quantities including 
the mass density and magnetic energy density. These QPOs result from 
instabilities at the jet-disk interface and strongly affect the jet dynamics.
The effects on the jet can clearly be seen in Figure 
\ref{fig:harm_data_thickdisk7} as density enhancements in the funnel region.

\begin{figure}
    \includegraphics[width=\linewidth]{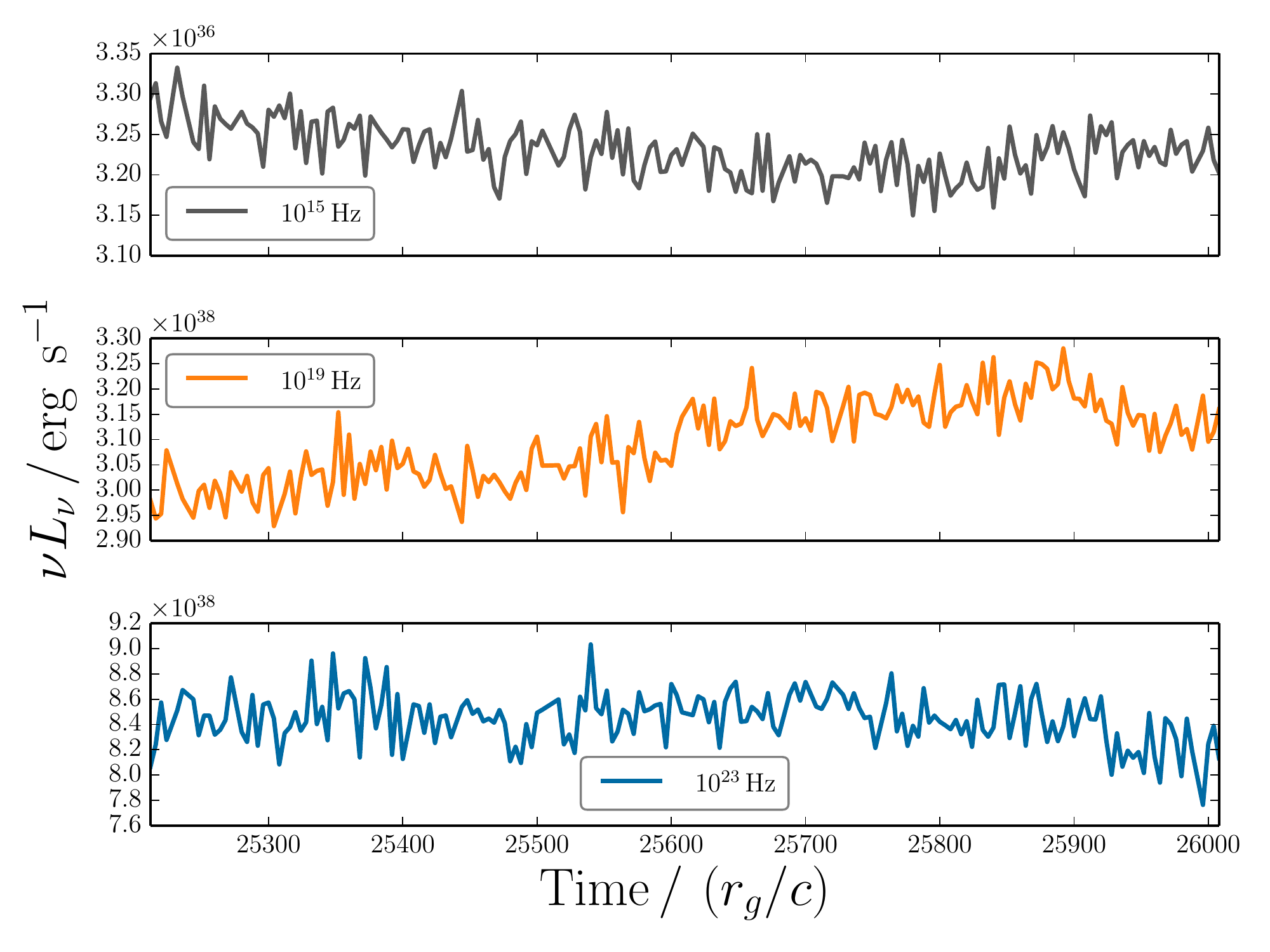}
    \caption{Lightcurves at $10^{15}\,\text{Hz}$, $10^{19}\,\text{Hz}$, and 
        $10^{23}\,\text{Hz}$ corresponding to synchrotron from the disk, 
        synchrotron from the jet, and scattering from the jet, respectively. 
        These were calculated with $\R_d = 10,\, \R_j = 3$.}
    \label{fig:qpos}
\end{figure}

\begin{figure}
    \includegraphics[width=\linewidth]{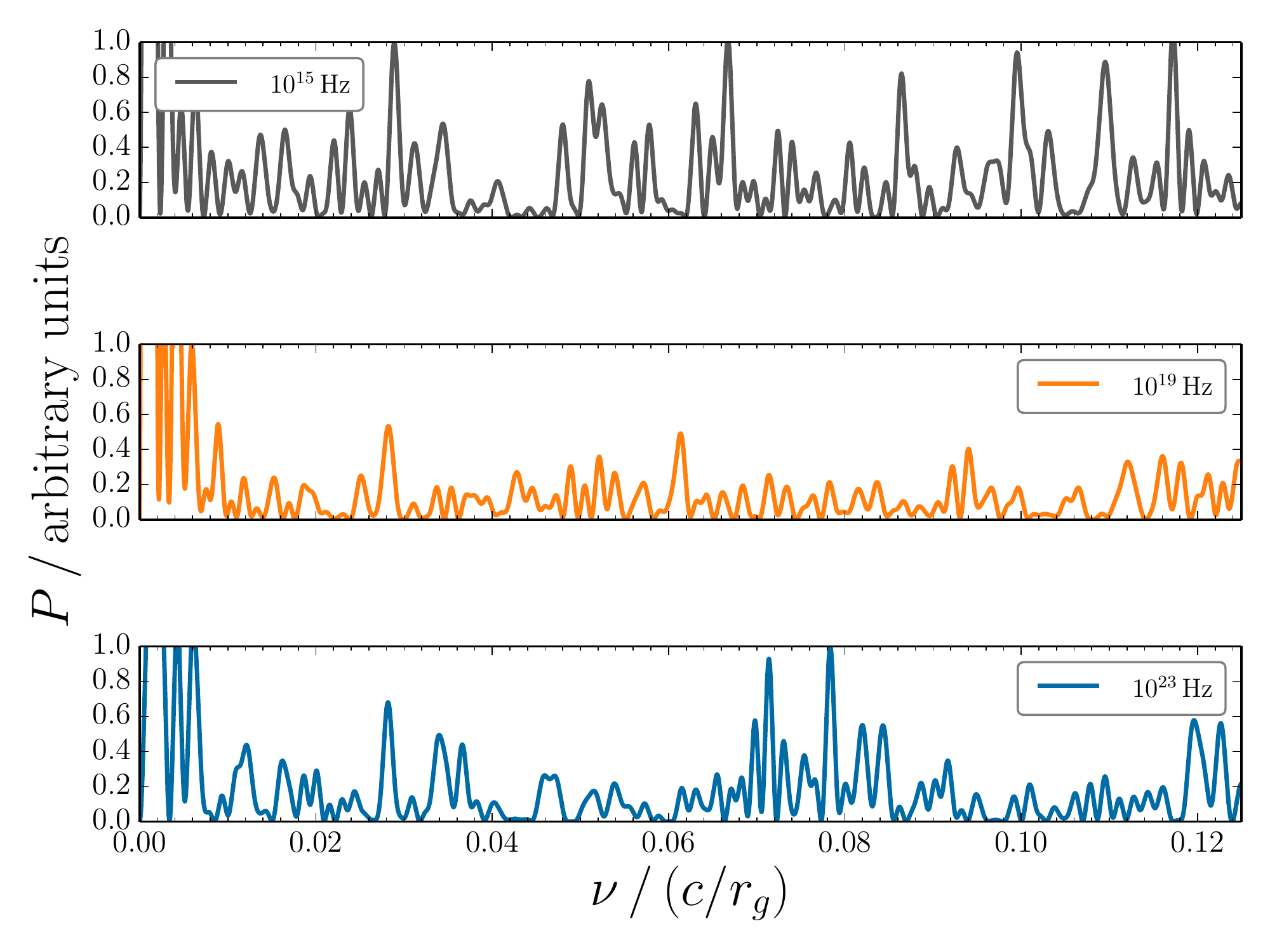}
    \caption{Power spectral densities for the lightcurves shown in Figure 
        \ref{fig:qpos}.}
    \label{fig:psds}
\end{figure}

\citet{SM13} tested the observability of the QPOs in the context of Sgr A* for
synchrotron emission at submillimeter wavelengths. 
In the present work, we investigate the detectability at higher frequencies in
the case of XRBs, and extend the previous analysis to include Comptonization.
The lightcurves in Figure \ref{fig:qpos} show variability at 
$10^{15}\,\text{Hz}$, $10^{19}\,\text{Hz}$, and $10^{23}\,\text{Hz}$, 
during a quasi-steady period of the MAD simulation 
(i.e., well after $t \approx 8000r_g/c)$. In Figure \ref{fig:psds}, we show the 
power spectral density of these curves. 
We find that the lightcurves are very noisy and show no clear QPO signal.
The lack of a clear QPO signal with Comptonization is an 
interesting result, and could have important implications 
for future efforts aimed at detecting QPOs at high frequencies.

\section{Summary and discussion}
\label{sec:discussion}
In this work, we calculated the spectrum of a RIAF in the context of the
low/hard state in XRBs, with the goal of identifying high-energy signatures of
jets in these systems. We investigated both MAD and non-MAD RIAFs, and find the 
following observational signatures of jet emission: 
(i) A significant peak in the $\gamma$-rays at $\sim 10^{23}\,\text{Hz}$.
(ii) A break in the optical/UV spectrum where it transitions from disk to jet 
dominated, changing from $\nu L_\nu \sim \nu^0$ at lower frequencies to 
$\nu L_\nu \sim \nu$ at higher frequencies. This is followed by a second break
around $\sim 10^{18}\,\text{Hz}$, where the spectrum roughly returns to 
$\nu L_\nu \sim \nu^0$, with ``wiggles'' in the luminosity of a factor of a few.
(iii) A pronounced peak near or below $\sim 10^{14}\,\text{Hz}$ indicates that 
jet emission contributes significantly to the X-rays.
These signatures are present across a range of proton-to-electron temperature 
ratios. 

Comparing the spectra in Figures \ref{fig:Rd_gtr_Rj} and 
\ref{fig:nonMAD_Mdot_10_3}, we find that spectra from our MAD model are 
almost completely jet dominated while those from our non-MAD model are dominated 
by the disk. In particular, the X-rays are produced by synchrotron self-Compton 
from the disk in our non-MAD model, while jet synchrotron emission dominates 
the X-rays in our MAD model.
Our results suggest that the two competing models of X-ray production in XRBs,
namely the synchrotron and synchrotron self-Compton models, are realised 
separately in MAD and non-MAD accretion flows, respectively.
Therefore, an investigation of the observational signatures of MAD vs non-MAD 
systems could provide valuable insights into breaking the degeneracy between 
these X-ray models. We will study these observational signatures further in a 
future work.

In our MAD model, 
we investigated the evolution of the jet and disk emission during a
large-scale magnetic field inversion in which the BZ jet is quenched and a new
transient jet is launched. 
This transient jet is qualitatively similar to those observed during
state transitions in XRBs \citep{Dexter+14}.
During the field inversion, the X-ray and $\gamma$-ray 
luminosities vary dramatically on a short timescale of $\sim 0.1$s.
The ratio of the $\gamma$-ray and X-ray luminosities changes from
$L_\gamma / L_X > 1$ in the steady BZ jet to $L_\gamma / L_X < 1$ during the 
transient outburst, and so is potentially an important observational signature 
of this process.
Furthermore, although outside the scope of the current work, we expect
to find significant variability in the radio at later times, as the hot plasmoid 
propagates outward and disrupts the flow at large radii.
Thus, a time lag between the fast correlated X-ray/$\gamma$-ray variability and 
radio variability could be a further indication of such a transient outburst.

The effects of QPOs on the jet dynamics were discussed in \citet{MTB12}, 
and their effects on disk emission were discussed in \citet{SM13}. 
Here, we extended this analysis to include the effects of Comptonization. 
Our results are noisy and show no clear QPO signal. 
This non-detection of the QPO is potentially important for future campaigns
aimed at detecting QPOs at high-frequencies.
The analysis here was carried out using a single electron temperature 
prescription, however, it is possible that different temperature prescriptions 
might reveal the QPO. We leave a more complete analysis of this jet-QPO 
variability to future work.

Our analysis was carried out for a limited range of fluid models and
temperature ratios, however, it is straightforward to estimate how the spectra 
would change with variations in $n$, $\Theta$, and $B$.
The synchrotron and inverse Compton peak frequencies scale with fluid properties
as $\nu_\text{syn} \sim \Theta^2 B$, and 
$\nu_\text{IC} \sim \Theta^2 \nu_\text{syn}$, respectively.
The heights of these peaks scale as
$(\nu j_\nu)_\text{syn} \sim n\Theta^2 B^2$, and 
$(\nu j_\nu)_\text{IC} \sim y (\nu j_\nu)_\text{syn}
\sim n\Theta^2 (\nu j_\nu)_\text{syn}$.
The saturation frequency is proportional to the electron temperature, 
$\nu_\text{sat} \sim \Theta$.
We can then scale our XRB results to AGN as follows.
Assuming that the accretion rate is proportional to 
the black hole mass,
the magnetic field, number density, and electron temperature in RIAFs vary with
$M$ as $B \sim M^{-1/2}$, $n \sim M^{-1}$, and $\Theta \sim M^0$
(see the discussion about scaling the \texttt{HARM} data to a particular system 
in Section \ref{subsec:rad transport}).
With these relationships, and the dependence of the spectral features on these
quantities as outlined above, we can scale our results to
arbitrary black hole masses. 

The most significant limitation of the current work is the assumption of a
thermal distribution of electrons. This may be a reasonable assumption for the
disk, however it is likely that the jet will contain a significant amount of
non-thermal particles due to shocks and magnetic reconnection. 
Also, the ``fast light'' approximation, which we use for computational 
efficiency, is an oversimplification since the dynamical time of the accretion 
disk and jet can be close to the light crossing time. 
We will extend this analysis to include the effects of non-thermal particles and
time-dependence in a future work.

\acknowledgments 
The authors would like to thank the anonymous referee
for many helpful suggestions that have improved the quality of the manuscript.
MOR is supported by the Irish Research Council by grant number GOIPG/2013/315.
This research was partially supported by the European Union Seventh Framework 
Programme (FP7/2007-2013) under grant agreement no 618499.
JCM acknowledges NASA/NSF/TCAN (NNX14AB46G), NSF/XSEDE/TACC (TGPHY120005), and 
NASA/Pleiades (SMD-14-5451).



\bibliographystyle{apj}
\bibliography{jet_signatures}

\end{document}